\documentclass[twocolumn,amssymb,nobibnotes,aps,showpacs,pra]{revtex4}
  \usepackage[dvips]{color}
  \usepackage{newlfont}
  \usepackage{graphicx}
  \usepackage{amssymb}
  \usepackage{amsmath}
  \usepackage[latin1]{inputenc}
  \usepackage{float}
  \usepackage{graphicx}
\newcommand{\br}{\begin{eqnarray}}
\newcommand{\er}{\end{eqnarray}}

\def\rg{\rangle}
\def\lg{\langle}
\begin{document}
\title{Unconditional Bell-type state generation for spatially separate
trapped ions}
\author{F. L. Semi\~ao}
\author{R. J. Missori}
\author{K. Furuya}
\affiliation{Instituto de Física `Gleb Wataghin', Universidade Estadual de
Campinas,\\ CP 6165, 13083-970, Campinas, SP, Brazil.}
\begin{abstract}
We propose a scheme for generation of maximally entangled states
involving internal electronic degrees of freedom of two distant
trapped ions, each of them located in a cavity. This is achieved by
using a single flying atom to distribute entanglement. For certain
specific interaction times, the proposed scheme leads to the
non-probabilistic generation of a perfect Bell-type state. At the
end of the protocol, the flying atom completely disentangles from
the rest of the system, leaving both ions in a Bell-type state.
Moreover, the scheme is insensitive to the cavity field state and
cavity losses. We also address the situation in which
dephasing and dissipation must be taken into account for the flying atom
on its way from one cavity to the other, and discuss the
applicability of the resulting noisy channel for performing quantum
teleportation.
\pacs{03.67.Mn, 42.50.Pq}
\end{abstract}

\maketitle

The idea of combining {\it stationary} and {\it flying} qubits in a
quantum network has recently attracted much interest from the
quantum information community \cite{lim,zoller}. In such a network,
the nodes are composed of stationary qubits (typically matter
qubits) and the information is carried between them by the flying
qubits, usually photons. This concept forms the base for an
alternative route to finding a scalable technology for building
quantum computers in a distributed way. Actually, the ability to
inter-convert stationary and flying qubits and also faithfully
transmit the latter between distant locations is part of what is
known as DiVincenzo requirements for the physical implementation of
quantum computation and information \cite{Divincenzo}. These
requirements seem to provide the necessary resources for any useful
use of quantum computers. In order to practically implement those
networks, distributed quantum systems in typical cavity-QED settings
have been considered in several papers
\cite{martin,cirac,paternostro,zou,barrett,serafini,cabrillo,feng,ou,outros}.
Proposals for the generation of entanglement between two spatially
separated matter qubits, without direct interaction, has often made
use of detection of another system, normally photons
\cite{cabrillo,feng,ou,outros}. Most of those schemes make use of
either $\Lambda$-type three-level or four level atoms/ions trapped
in distant cavities where the entanglement between them is
established via interference induced effects. Bose et. al
\cite{martin} and Cabrillo et. al. \cite{cabrillo} independently
made the seminal proposals. In Bose et. al scheme \cite{martin},
atoms inside lossy cavities become entangled by Bell measurements on
photons escaping from the cavities (achieved with the use of
beam splitters) and entanglement swapping. In Cabrillo et. al.
proposal \cite{cabrillo}, entanglement is created by driving the
atoms with a laser pulse and detecting subsequent spontaneous
emission; Feng et.al. \cite{feng} proposed a scheme using
interference of photons leaking out the cavities, and a
generalization for $3$-qubits and GHZ/W state generation has been
proposed by Ou et. al. \cite{ou}. Recently Barrett and Kok
\cite{barrett} used Bose et. al scheme \cite{martin} to propose
scalable distributed QC with nondeterministic entangling operations,
and Lim {\sl et al} \cite{lim} proposed a repeat-until-success gate
operation allowing to eventually perform it in a deterministic
fashion. Despite the evident importance and applicability of such
probabilistic and repeat-until-success schemes, it is always
desirable to have a non-probabilistic way of generating entanglement
and related gate operations. In this context, Clark {\sl et al}
\cite{clark} have proposed a scheme for unconditional preparation of
entanglement between atoms trapped in separate cavities by employing
quantum reservoir engineering in a appropriate cascade cavity-QED
setting.

Here, we propose a new unconditional method for generating the
maximally entangled state (Bell-type state)
\begin{eqnarray}
|\Psi_{1,2}\rg=\frac{1}{\sqrt{2}}(|e_{1},g_{2}\rg +
|g_{1},e_{2}\rg), \label{bell}
\end{eqnarray}
between two distant two-level trapped ions ($1$ and $2$), where
$|e_i\rangle$ and $|g_i\rangle$ stand for the electronic excited and
ground state of the ion $i$. This non-probabilistic generation
protocol makes use of a flying two-level atom sequentially
interacting with both ions and promoting the establishment of
entanglement between them. The current cavity-QED experiments
already manipulate flying atoms with high degree control and employ
it to implement quantum dynamics \cite{haroche}. On the other hand,
the trapping and local laser manipulation of ions has also improved
and many important experiments have been performed \cite{ions}.
Therefore, thinking of an union of both settings seems to be a very
promising idea and it is the very core of our proposal. Although
there are many proposals for quantum state manipulation or quantum
computing using trapped ions in cavities \cite{cavion}, our scheme
seems to be the first one based on the interaction of flying atoms
with trapped ions.

In this letter, we show how the state (\ref{bell}) may be perfectly
generated in an ideal case, and then we include dissipation and dephasing for the {\it flying
qubit} (atom) during its course between the cavities containing
\emph{stationary qubits} (ions). The establishment of entanglement
between distant parties forms a quantum channel which in association
with classical communication may be used for several applications
such as superdense coding \cite{bennett1} and quantum teleportation
\cite{bennett2}. Thus, we evaluate the fully entangled fraction
\cite{fef} which is directly related to the fidelity of those
applications \cite{fid,fid2}.

\begin{figure}[h]
\centering
\includegraphics[width=0.7\columnwidth]{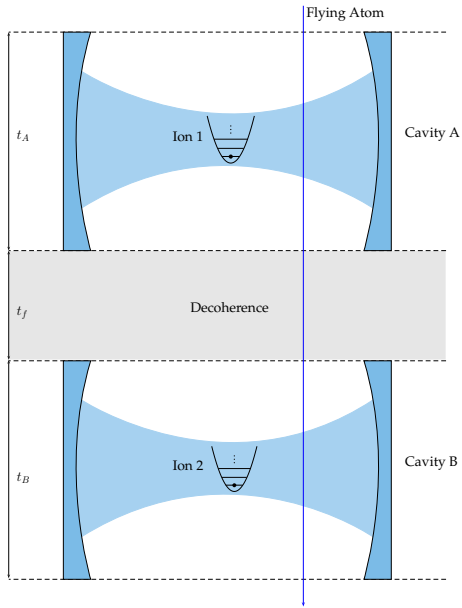}
\caption{\label{fig1}A sketch of the system comprising two-level
trapped ions inserted in spatially separated cavities and a flying
atom sent to cross both cavities. The flying atom may experience
decoherence due to amplitude and phase damping when crossing the
region between the cavities.}
\end{figure}

The system under consideration consists of two distant cavities $A$
and $B$, each of them containing one trapped ion inside, $1$ and
$2$, and a flying two-level atom crossing both cavities, as shown
schematically in Figure \ref{fig1}.

The Hamiltonian describing the interaction of a trapped ion, the
flying atom and the cavity field in one of the cavities is
\begin{eqnarray}
{\hat{H}}={\hat{H}}_{0}+{\hat{H}}_{int},
\end{eqnarray}
with
\begin{eqnarray}
{\hat{H}}_{0}=\hbar\nu{\hat{a}}^{\dag}{\hat{a}}+
\hbar\omega{\hat{b}}^{\dag}{\hat{b}}+ \frac{1}{2}\hbar\omega_{1(2)}
\hat{\sigma}_{z}^{1(2)}+ \frac{1}{2}\hbar\omega_{f}
\hat{\sigma}_{z}^{f},
\end{eqnarray}
and
\begin{eqnarray}
{\hat{H}}_{int}&=& \hbar
g_{f_{A(B)}}(\hat{\sigma}_{+}^{f}{\hat{b}}+\hat{\sigma}_{-}^{f}{\hat{b}}^{\dag})\nonumber
\\ &&+\hbar
g_{1(2)}\cos{[\eta({\hat{a}}^{\dag}+{\hat{a}})]} 
(\hat{\sigma}_{+}^{1(2)}{\hat{b}}+\hat{\sigma}_{-}^{1(2)}{\hat{b}}^{\dag}),
\end{eqnarray}
\noindent where the indexes $1$, $2$, and $f$ are used to label the
respective ions, and the flying atom, $\hat{\sigma}$'s are the
Pauli atomic operators, $g_{1(f_A)}$ is the coupling constant
between the trapped ion $1$ (flying atom) and the field mode in cavity A, $g_{2(f_B)}$ is the coupling constant
between the trapped ion $2$ (flying atom) and the field mode in cavity B,
$\omega_{1(2)[f]}$ is the atomic transition frequency of the ion
$1(2)$[flying atom], ${\hat{b}^{\dag}}$$({\hat{b}})$ is the creation
(annihilation) operator of the cavity mode (frequency $\omega$),
${\hat{a}^{\dag}}({\hat{a}})$ is the bosonic operator of the ion
vibrational mode (frequency $\nu$), and $\eta=2\pi a_0/\lambda_0$ is
the Lamb-Dicke parameter, being $a_0$ the rms fluctuation of the
ion's position in the lowest trap eigenstate, and $\lambda_0$ is the
cavity field wavelength.

We are now going to use an effective Hamiltonian in the case which
the trapped ions and the flying atom are kept far off-resonance with
the field ($\Delta=\omega-\omega_{1(2)[f]}\gg g_{1(2)[f_{A(B)}]}$). In this
case, there is no energy exchange between the cavity field and the
matter qubits. If one also chooses the frequencies of the system
such that no sidebands are to be excited \cite{kerr}, i.e
$\Delta\ll \nu$, and maintain the resonance
between ions and flying atom $(\omega_1=\omega_2=\omega_f)$, it is
possible to end up with the following Hamiltonian in the interaction
picture
\begin{eqnarray}
{\hat{H}}_{int}^{1(2)}=\lambda_{1(2)}(\hat{\sigma}_{+}^{f}\hat{
\sigma}_{-}^{1(2)}
+ \hat{\sigma}_{-}^{f}\hat{\sigma}_{+}^{1(2)}) ,\label{h}
\end{eqnarray}
\noindent where ${\lambda}_{1(2)} = g_{1(2)} g_{f_{A(B)}}/\Delta$ is the
effective coupling constant in cavity A(B). 

Now, it is important to consider a specific example of ion and neutral atom species that meet the resonance conditions assumed in the derivation of Hamiltonian (\ref{h}). First, in order to prevent the qubits to decohere during the process, long-lived transitions are needed for both the ions and the neutral atom. From the ion side, this could be achieved by using ions with a ground state that presents hiperfine structure. In this case, the application of magnetic fields can split the hyperfine levels to a few tens of Gigahertz. On the other hand, long-lived transitions of a neutral atom can be chosen to be circular Rydberg states, also separated by tens of Gigahertz. Flying Rb atoms prepared in circular Rydberg states and interacting with microwave photons have been used in a series of entanglement experiments \cite{haroche}.  It is then clear that in order to manipulate such ionic and atomic states, microwave radiation is a natural choice. This fact seems to spoil the scheme because the Lamb-Dicke parameter for the ions would be terribly small for the ordinary traps. However, recent experiments have made use of a modified ion trap in which a modified effective Lamb-Dicke parameter can be made sufficiently large for our proposals \cite{Yb1}. Actually, for Yb$^+$ ions, all quantum operations involving motion and internal degrees-of-freedom usually carried out with optical fields can now be implemented using microwave radiation. 

We propose the use of Yb$^+$ ions and flying Rb atoms prepared in circular Rydberg states in the implementation of the ideas suggested in this paper. The reported quantum optical experiments with $^{171}$Yb$^+$ ions have used the $S_{1/2}$ ground-state hyperfine doublet $\{|g\rangle\equiv|S_{1/2},F=0\rangle, |e\rangle\equiv|S_{1/2},F=1,m_F=0\rangle\}$ as the qubit, and we make this choice here. Due to the application of static magnetic fields, the qubit transition used in \cite{Yb1} is about $12.6$GHz. Another hyperfine states could also be chosen as shown in \cite{Yb2}. The qubit transition in circular Rydberg states of Rb is reported to be $51.1$Ghz. We feel that a moderate increasing in the strength of the magnetic field acting on Yb$^+$ could lead to the resonance $(\omega_1=\omega_2=\omega_f)$, as assumed in the derivation of (\ref{h}). We should still mention that miniature Paul traps consisting of a ring electrode of diameter 2mm and endcaps electrodes spaced $\sqrt{2}$mm apart have been used for performing some experiments involving Yb$^+$ ions \cite{Yb1}. This facilitates the insertion of the trap in the microwave cavity whose mirrors are spaced 27mm apart in the quantum optics experiments involving flying Rb atoms.

The final form (\ref{h}) is essentially the Hamiltonian used by Zeng and Guo in \cite{zheng}, but 
with the advantage that one of the qubits is trapped and kept within the cavity allowing further manipulation with relative ease. This particularly simple form of the Hamiltonian (\ref{h}) has been
obtained by also considering that both ions are initially cooled
down to their lowest trap state. Otherwise, the motion would couple
to the electronic degrees of freedom. Since the Hamiltonian
(\ref{h}) does not depend on the cavity field, the Rabi frequency
will be simply $\lambda_{1(2)}$ in A(B). This fact makes our scheme
insensitive to the cavity field state and cavity decay
\cite{zheng,sorensen}. It would work even for a thermal field with a
few photons. Now, consider that both ions are initially in their
electronic ground state and the flying atom is in its excited state,
i.e $|\psi(0)\rangle=|e_{f},g_{1},g_{2}\rg$. According to (\ref{h}),
if the flying atom takes {\bf $\lambda_1 t_A=\pi/4$} to cross cavity
A, the state of the system will be
\begin{equation}
|\psi(t_A)\rangle=\frac{1}{\sqrt{2}}(|e_{f},g_{1}\rg -i
|g_{f},e_{1}\rg)\otimes|g_2\rangle.\label{psiA}
\end{equation}
Then, the atom is let to fly from one cavity to the other and it
takes a time of flight $t_f$. In the ideal case (without
decoherence), the evolution of the three subsystems during the
interval $t_f$ is local, and unitary, leading to no change in the
entanglement shared between them. However, if losses or dephasing
are included, the entanglement shared between them will change due
to the coupling with the environment. We will deal with these
aspects later on in this paper. Now, the flying atom reaches cavity
B, and after spending {\bf $\lambda_2 t_B=\pi/2$} to cross it, the
global state of the system will be (except for a global phase)
\begin{equation}
|\psi\rangle=|g_{f}\rangle\otimes|\Psi_{1,2}\rg,\label{gbell}
\end{equation}
where $|\Psi_{1,2}\rg$ is the Bell-type state (\ref{bell}).  We can
see from (\ref{gbell}) that at the end of the protocol, the state of
the flying atom completely factorizes from the rest of the system
which is left in a perfect maximally entangled state
$|\Psi_{1,2}\rg$. Therefore, the scheme proposed here does not rely
on probabilistic outcomes of any measurement process, thus being an
unconditional generation protocol. It is important to notice that the protocol requires different interaction times at the two cavities, and since the velocity of the flying atom is supposed to not vary, just the coupling constants $\lambda_1$ and $\lambda_2$ can be adjusted. Considering the same species of ions and the same field frequency $\omega_f$ in both cavities, the only free parameter to control the interaction times is the volume of the cavities which enters in the definition of $g_{f_{A(B)}}$ and $g_{1(2)}$ \cite{dephasing}.

In real experiments, the flying atom on its way from one cavity to
the other may collide with other atoms or molecules resulting in
dephasing \cite{dephasing}. Also, depending on how far the cavities
are set from each other, the flying atom may spontaneously decay due
to the coupling with the electromagnetic modes of free space. Once the
cavity decay does not destructively affect our scheme, as explained
before, phase and amplitude damping of the flying atom seem to be
the most important source of loss of quantum coherence here. We are
now going to model such noise mechanisms by the standard method of
master equations. The system master equation in the interaction
picture describing spontaneous emission and phase damping of the
flying atom on its way from cavity A to cavity B is given by
\begin{eqnarray}
\frac{\partial\hat{\rho}(t)}{\partial t} &=&
\frac{\gamma}{2}\left[2\hat{\sigma}_{-}^{f}\hat{\rho}(t)\hat{\sigma}_{+}^{f}-
\hat{\sigma}_{+}^{f}\hat{\sigma}_{-}^{f}
\hat{\rho}(t)
-\hat{\rho}(t)\hat{\sigma}_{+}^{f}\hat{\sigma}_{-}^{f}\right]\nonumber\\
 &&+
\frac{\gamma_{p}}{2}\left[\hat{\sigma}_{z}^{f}\hat{\rho}(t)
\hat{\sigma}_{z}^{f}-\hat{\rho}(t)\right],
\label{master}
\end{eqnarray}
where $\gamma(\gamma_{p})$ is the atomic decay (dephasing) rate.
This equation is to be solved with the initial condition
$\hat{\rho}(0)=|\psi(t_A)\rangle\langle \psi(t_A)|$, where
$|\psi(t_A)\rangle$ is the global state (\ref{psiA}) after the
flying atom has left the cavity A. The solution of the master
equation for such a initial condition is
\begin{eqnarray}
\hat{\rho}(t_{f}) &=& \frac{1}{2}\bigl[|g_f,e_{1}\rg\lg g_f,e_1|-i
e^{-(\gamma+\gamma_{p})t_{f}}|g_f,e_1\rg\lg
e_f,g_1|\nonumber \\
&&+ i e^{-(\gamma+\gamma_{p})t_{f}}|e_f,g_1\rg\lg g_f,e_1|
\nonumber \\
&&+(1-e^{-\gamma t_{f}})|g_{f},g_{1}\rg\lg
g_{f},g_{1}|\bigr]\otimes|g_2\rangle\langle g_2|,\label{rho0}
\end{eqnarray}
where $t_f$ is the time of flight. Now, the flying atom reaches
cavity B, and there it follows a unitary evolution according
to (\ref{h}). Considering the initial state of the system to be
(\ref{rho0}), and just like in the ideal case the atom still spends
{\bf $\lambda_2 t_B=\pi/2$}, one will obtain after tracing out the
flying atom
\begin{eqnarray}
\hat{\rho}_{1,2} &=& \frac{1}{2}\bigl[|e_{1},g_{2}\rg\lg
e_{1},g_{2}|+ e^{-(\gamma+\gamma_{p})t_{f}}|e_{1},g_{2}\rg\lg
g_{1},e_{2}|+\nonumber \\
&&+ e^{-(\gamma+\gamma_{p})t_{f}}|g_{1},e_{2}\rg\lg
e_{1},g_{2}|+e^{-\gamma t_{f}}|g_{1},e_{2}\rg\lg g_{1},e_{2}|
\nonumber \\
&&+(1-e^{-\gamma t_{f}})|g_{1},g_{2}\rg\lg
g_{1},g_{2}|\bigr]\label{rho}.
\end{eqnarray}
The state (\ref{rho0}) involving the spatially separated trapped
ions is the result of our generation protocol when the flying atom
employed to distribute entanglement is affected by amplitude and
phase damping. The first thing to be noted is that the state
(\ref{rho}) has not as much entanglement as the Bell-type state
generated in the ideal case. Actually, the entanglement in
(\ref{rho}) decreases exponentially with the time of flight as
measured by the concurrence $C$ \cite{conc}, which for (\ref{rho})
is given by $C(t_f)=e^{-(\gamma+\gamma_p)t_f}$. If Alice and Bob
share a two-qubit mixed state such as (\ref{rho}), they can try to
teleport the unknown state of a third qubit using local quantum
operations and classical communication (LQCC). It has been
demonstrated \cite{fid} that the optimal fidelity of teleportation
$f_{max}$ attainable using LQCC is connected to a quantity called
fully entangled fraction $F_{max}$ which is defined as
$F_{max}=\max_{|\Psi\rangle}\langle \Psi|\hat{\rho}_{1,2}
|\Psi\rangle$. The maximization here is taken over all maximally
entangled states $|\Psi\rangle $, i.e all states that can be
obtained from a singlet using local unitary transformations. The 
relation between both quantities is $f_{max}=(2F_{max}+1)/3$  
\cite{fid}. The fully entangled fraction may be readily evaluated 
writing $\hat{\rho}_{1,2}$ in a suitable basis and finding the 
eigenvalues of its real part as suggested in \cite{fef,fid2}. 
Following their receipt it is not difficult to see that for the 
noisy channel $\hat{\rho}_{1,2}$ given by (\ref{rho}) the fully 
entangled fraction reads
\begin{equation}
F_{max}=\frac{1}{4} (1+ e^{-\gamma t_f} +2 e^{-(\gamma+\gamma_{p})
t_f}).\label{fef}
\end{equation}
A classical channel can give at most a fidelity equal to $2/3$ that
is achieved when Alice simply performs a measurement on the unknown
qubit and tells Bob the outcome \cite{popescu1}. It follows that in
order to gain a real improvement coming from quantum mechanics one
must have $F_{max}\geq 1/2$. For special cases, we can get simple
but very important upper limits for the time of flight $t_f$. For
the pure amplitude damping channel, for instance, one must have
$t_f\leq\ln(3)\gamma^{-1}$, and for equal dephasing and damping
($\gamma_p=\gamma$) channels, $t_f\leq\ln(2)\gamma^{-1}$. Different
choices of $\gamma$ and $\gamma_p$ lead to other maximal values of
$t_f$ obtained from (\ref{fef}). For the flying Rydberg atom with principal quantum number
$n \approx 50$ having $\gamma \approx 2 \times 10^2 s^{-1}$
\cite{zheng}, it will lead to $t_f^{max}=5$ ms for the pure
amplitude damping case. Considering the typical velocity of the
flying Rydberg atom to be $v \approx 3 \times 10^2 m/s$
\cite{haroche,zheng}, the maximal distance between the cavities and
so the distance between the entangled pair of ions should be around
1.5m. This is clearly just a theoretical limit and we do not expect that it might be achieved so easily. There are many technicalities not included in the calculation of such distance.

We should now critically comment on the weakness and usefulness of our proposal. It relies heavily on the ability to produce and to move atoms with very well-controlled velocities. Otherwise, the scheme would not be deterministic for it supposes an ideal situation where the velocity of the flying atom is precisely specified. The state-of-art in the manipulation of flying Rydberg atoms has achieved velocity selection with a width of $\pm 0.4$m/s \cite{doppler}. This is enough level of control for our proposal to be, at some high degree, and in relation to this aspect, deterministic i.e. with a small probability of error. This high degree of control over the velocity of the flying atom is achieved by the use of Doppler-selective optical pumping techniques \cite{doppler}. It is also important to be sure that just one atom is crossing the cavity at a time. It is reported that the circular states excitation process prepares one atom per pulse with a probability as high as 90\% \cite{hreview}. The other events are rejected when they take place. 

Although our scheme is theoretically deterministic, it would in practice have to be dealt with such probabilistic experimental aspects. Actually, it is not a problem in itself since this is the case in any practical implementation of a theoretical proposal. It is also important to mention that the present proposal is not intended to be useful for large distance distributed computation. It is a common belief that this is more suitable for photonic channels. On the contrary, our proposal aims to provide a generation method of entangled states involving macroscopically separated ions to be used, for instance, for fundamental experiments as quantum teleportation. Macroscopic distance means a separation much longer than the extent of the wave function of each ion. A few millimeters or centimeters apart must be considered a macroscopic distance. Another several limitation for large distance distributed computation using flying atoms is the demand for large vacuum chambers. However, for a separation of a few centimeters, only one small vacuum chamber would be needed to implement the scheme. Having in mind the scope of application of our proposal, it can be an alternative for the repeat-until-success schemes when dealing with short but macroscopic distances. On the other hand, the probabilistic schemes involving photonic flying qubits are more appropriate for application in tasks involving large distances.

In summary, we have shown a new scheme to generate entangled states
involving distant parties. To our knowledge, it is the first
proposal based upon the interaction of flying atoms with trapped
ions. Even though the cavity is used to induce an indirect interaction
between them, our scheme is robust against cavity losses and
insensitive to the cavity field state. In this case, our scheme
results in the unconditional generation of a perfect Bell-type
state. We have also considered a situation where spontaneous atomic
decay and dephasing are included. For this setting, we
have discussed limits of the applicability of the generated state
for performing quantum teleportation and discussed aspects of a 
practical implementation.

\acknowledgments{
FLS would like to thank R. Blatt for valuable discussions; RJM
acknowledges financial support from CAPES, FLS from FAPESP under
project $05/04533-7$ and KF to CNPq (partial) under projects no. $303662/2004-2$ and $420248/2005-6$.}


\begin{thebibliography}{99}
\bibitem{lim} J. I. Cirac {\it{et al}}, Phys. Rev. A {\bf 59} 4249 (1999); Y.L. Lim
{\it{et al}}, Phys. Rev. Lett. {\bf 95} 030505 (2005); Y.L. Lim
{\it{et al}}, Phys. Rev. A {\bf 73}, 012304 (2006).
\bibitem{zoller}W. Yao {\it{et al}}, Phys. Rev. Lett. \textbf{95}, 030504
(2005); L.-m. Liang and C.-z. Li, Phys. Rev. A \textbf{72}, 024303
(2005); S.-L. Zhu {\it{et al}},  {\it{ibid}} \textbf{68}, 034303
(2003).
\bibitem{Divincenzo}D. P. Divincenzo, Fortschr. Phys. {\bf{48}}, 771
(2000).
\bibitem{martin} S. Bose, P.Knight, M.B. Plenio and V. Vedral,
Phys. Rev. Lett.  \textbf{83}, 5158 (1999).
\bibitem{cabrillo} C. Cabrillo {\it{et al}}, Phys. Rev. A {\bf{59}}, 1025
(1999).
\bibitem{cirac} J.
I. Cirac  {\it{et al}}, Phys. Rev. Lett. \textbf{78}, 3221 (1997).
\bibitem{paternostro} M. Paternostro {\it{et al}}, J. Mod. Opt.
{\bf 50} $2075$ ($2003$).
\bibitem{zou} X. B. Zou and W. Mathis, Phys. Rev. A {\bf 71}, 042334 (2005).
\bibitem{barrett} S. D. Barrett and P. Kok,  Phys. Rev. A {\bf 71} 060310
(2005).
\bibitem{serafini} A. Serafini {\it{et al}}, Phys. Rev. Lett. {\bf 96},
010503 (2006).
\bibitem{feng} X. -L. Feng {\it{et al}}, Phys. Rev. Lett. {\bf{90}},
217902 (2003).
\bibitem{ou} Y.-C. Ou {\it{et al}}, J. Phys. B: At. Mol. Opt. Phys.
{\bf 39}, 7 (2006).
\bibitem{outros} D. E. Browne {\it{et al}}, Phys. Rev. Lett.
{\bf{91}}, 067901 (2003); C. Simon and W. T. M. Irvine, {\it ibid.}
{\bf{91}} 110405 (2003); L.-M. Duan and H. J. Kimble {\it ibid.}
{\bf{90}} 253601 (2003); X. B. Zou {\it{et al}}, Phys. Rev. A {\bf
69}, 052314 (2004); L.-M. Duan {\it{et al}}, {\it ibid.} {\bf 72},
032333 (2005).
\bibitem{clark} S. Clark {\it{et al}}, Phys. Rev. Lett. {\bf 91}, 177901
(2003).
\bibitem{haroche}M. Brune {\it{et al}}, Phys. Rev. Lett.
\textbf{76}, 1800 (1996); E. Hagley {\it{et al}}, {\it ibid.}
\textbf{79}, 1 (1997); G. Nogues {\it{et al}}, Nature (London)
\textbf{400}, 239 (1999); A. Rauschenbeutel {\it{et al}}, Phys. Rev.
Lett.
\textbf{83}, 5166 (1999); S. Osnaghi {\it{et al}}, {\it ibid.}
\textbf{87}, 037902 (2001); P. Bertet {\it{et al}}, {\it ibid.}
\textbf{89}, 200402 (2002); P. Bertet {\it{et al}}, {\it ibid.}
\textbf{88}, 143601 (2002); A. Auffeves {\it{et al}}, {\it ibid.}
\textbf{91}, 230405 (2003); T. Meunier {\it{et al}}, {\it ibid.}
\textbf{94}, 010401 (2005).
\bibitem{ions} C. Monroe  {\it{et al}}, {\it ibid.} \textbf{75},
4714 (1995); D. M. Meekhof  {\it{et al}}, {\it ibid.} \textbf{76},
1796 (1996); D. Leibfried  {\it{et al}}, {\it ibid.} \textbf{77},
4281 (1996); Ch. Roos  {\it{et al}}, {\it ibid.} \textbf{83}, 4713
(1999); M. Riebe  {\it{et al}}, Nature (London) \textbf{429}, 734
(2004); M. D. Barrett  {\it{et al}}, {\it ibid.} \textbf{429}, 737
(2004); C. F. Roos  {\it{et al}}, Phys. Rev. Lett {\bf 92}, 220402
(2004); P. C. Haljan {\it{et al}}, {\it ibid.} \textbf{94}, 153602
(2005); M. J. Madsen {\it{et al}}, {\it ibid.} \textbf{97}, 040505
(2006).
\bibitem{cavion}H. Zeng and F. Lin, Phys. Rev. A
\textbf{50}, 3589(R) (1994); V. Bu\v{z}ek {\it{et al}}, {\it ibid.}
\textbf{56}, 2352 (1998); A. C. Doherty {\it{et al}}, {\it ibid.}
\textbf{57}, 4804 (1998); A. S. Parkins and E. Larsabal, {\it ibid.}
\textbf{63}, 012304 (2000); F. L. Semi{\~a}o {\it{et al}}, {\it
ibid.} \textbf{64}, 024305 (2001); Xu Bo Zou {\it{et al}}, {\it
ibid.} \textbf{65}, 064303 (2002); C. Di Fidio {\it{et al}}, {\it
ibid.} \textbf{65}, 033825 (2002);R. Rangel {\it{et al}}, {\it
ibid.} \textbf{69}, 023805 (2004); G.-X. Li {\it{et al}}, {\it
ibid.} \textbf{71}, 063817 (2005).
\bibitem{bennett1} C. H. Bennett and S. J. Wiesner,  Phys. Rev. Lett. {\bf
69}, 2881 (1992).
\bibitem{bennett2} C. H. Bennett {\it{et al}}, Phys. Rev. Lett. {\bf 70},
 1895 (1993).
\bibitem{fef} C. H. Bennett {\it{et al}},
Phys. Rev. A \textbf{54}, 3824 (1996);
\bibitem{fid} M. Horodecki {\it{et al}}, Phys. Rev.
 A \textbf{60}, 1888 (1999)
\bibitem{fid2} J. Grondalski {\it{et al}}, Phys. Lett. {\bf A}
300, 573 (2002).
\bibitem{kerr} F. L. Semi\~ao and A. Vidiella-Barranco, Phys. Rev. A
{\bf 72}, 064305 (2005).
\bibitem{Yb1} F. Mintert and C. Wunderlich, Phys. Rev. Lett. {\bf 87}, 257904 (2001);  C. Wunderlich, C. Balzer, T. Hannemann, F. Mintert, W. Neuhauser, D. Rei\ss, and P. E. Toschek, J. Phys. B: At. Mol. Opt. Phys. {\bf 36}, 1063 (2003) .
\bibitem{Yb2} D. Mc Hugh and T. Twamley, Phys. Rev. A {\bf 71}, 012315 (2005).
\bibitem{zheng} S.-B. Zheng and G.-C. Guo, Phys. Rev. Lett. {\bf 85}, 2392
(2000); S. Osnaghi,  {\it{et al}}, Phys. Rev. Lett. {\bf 87},
037902 (2001).
\bibitem{sorensen}A. S{\o}rensen and K. M{\o}lmer, Phys. Rev. Lett. {\bf 82},
1971 (1999).
\bibitem{dephasing} H. J. Carmichael, {\it An open systems approach to quantum optics} (Springer-Verlag,
1993); D. F. Walls and G. J. Milburn, {\it Quantum Optics}
(Springer-Verlag, 1994); L. -M. Kuang, X. Chen, G. -H Chen, and M. -L Ge, Phys. Rev. A {\bf 56}, 3139
(1997).
\bibitem{conc} W. K. Wooters, Phys. Rev. Lett \textbf{80}, 2245 (1998).
\bibitem{popescu1}S. Popescu, Phys. Rev. Lett {\bf 72}, 797 (1994); S. Massar and S. Popescu;
Phys. Rev. Lett {\bf 74}, 1259 (1995); J. Preskill, \emph{Lecture
Notes} available at http://www.theory.caltech.edu/~preskill/ph229/.
\bibitem{doppler} E. Hagley, X. Ma\^itre, G. Nogues, M. Brune, J. M. Raimond, and S. Haroche, Phys. Rev. Lett. {\bf 79}, 1 (1997).
\bibitem{hreview} J. M. Raimond, M. Brune, and S. Haroche, Rev. Mod. Phys. {\bf 73}, 565 (2001).
(2002).
\end{thebibliography}
\end{document}